\documentstyle[psfig,epsfig,amssymb,a4paper]{jpconf}

\begin{document}

\title{ 
Critical behavior of two-dimensional fully frustrated XY systems}

\author{Martin Hasenbusch,$^{1}$ Andrea Pelissetto,$^2$ \footnote[3]{
Presented at the Conference ``Counting Complexity",
July 10--15, 2005, Dunk Island, QLD, Australia.
In celebration  of Tony Guttmann's 60th birthday.} Ettore
Vicari$\,^1$ } 

\address{$^1$ Dip. Fisica dell'Universit\`a di Pisa and
INFN, Largo Pontecorvo 2, I-56127 Pisa, Italy} 

\address{$^2$
Dip. Fisica dell'Universit\`a di Roma ``La Sapienza" and INFN, \\ P.le
Moro 2, I-00185 Roma, Italy} 

\ead{{\tt
Martin.Hasenbusch@df.unipi.it,} {\tt Andrea.Pelissetto@roma1.infn.it},
{\tt Ettore.Vicari@df.unipi.it} }

\date{\today}

\begin{abstract}
We study the phase diagram of the two-dimensional fully frustrated XY
model (FFXY) and of two related models, a lattice discretization of
the Landau-Ginzburg-Wilson Hamiltonian for the critical modes of the
FFXY model, and a coupled Ising-XY model.  We present Monte Carlo
simulations on square lattices $L\times L$, $L\lesssim 10^3$.  We show
that the low-temperature phase of these models is controlled by the same
line of Gaussian fixed points as in the standard XY model. We find that, 
if a model undergoes a unique transition by varying temperature,
then the transition is of first order. In the opposite case
we observe two very close transitions: a transition associated 
with the spin degrees of freedom and, as temperature increases, 
a transition where chiral modes become critical. If they are continuous,
they belong to the Kosterlitz-Thouless and to the Ising universality class,
respectively.  Ising and  Kosterlitz-Thouless behavior is observed
only after a preasymptotic regime, which is universal to some extent.
In the chiral case, the approach is nonmonotonic for most observables, 
and there is a wide region in which finite-size scaling is controlled by an 
effective exponent $\nu_{\rm eff} \approx 0.8$.  This explains the result
$\nu\approx 0.8$ of many previous studies using smaller lattices.
\end{abstract}



\section{Fully frustrated systems}
\label{sec1}

In the last few decades there has been a considerable interest in the 
consequences of frustration on the critical behavior of statistical systems.
The simplest example is the antiferromagnetic Ising model on a triangular 
lattice, whose Hamiltonian is
\begin{equation}
{\cal H} = J \sum_{\langle ij \rangle} \sigma_i \sigma_j,
\end{equation}
where $J$ is positive, $\sigma_i = \pm 1$, and the sum is extended
over all lattice nearest neighbors $\langle ij \rangle$. 
\begin{figure}[tb]
\includegraphics[width=17pc]{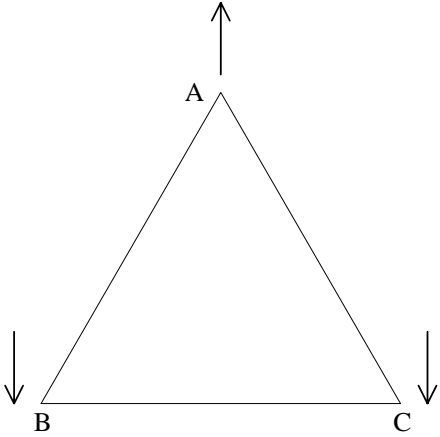}\hspace{2pc}%
\begin{minipage}[b]{17pc}
\caption{\label{frustration}
A frustrated triangle. If $\sigma_A = 1$, the local energy associated 
with links AB and AC is 
minimized by taking $\sigma_B = \sigma_C = -1$. The local energy
on link BC corresponds to a maximum: link BC is {\it frustrated}. 
}
\vspace{10mm}
\end{minipage}
\end{figure}
For $T\to 0$ the system tends to be antiferromagnetically ordered, i.e.
spins on nearest-neighbor sites prefer to be oppositely aligned.
However, this is not possible everywhere. For instance, see 
Fig.~\ref{frustration}, on any lattice triangle one link must 
be {\em frustrated},
i.e.~spins on the corresponding sites must be parallel so that the local
energy assumes its {\it maximum} value.
The presence of frustration has an important consequence. At variance with 
the ferromagnetic Ising model the antiferromagnetic one is 
disordered at any temperature: the large entropy forbids an ordering transition
\cite{Wannier-50,Houtappel-50}.\footnote{It is interesting to note that 
this is not true for the spin-$S$ antiferromagnetic Ising model if $S$ is 
large enough; see \cite{NMH-93,LHL-95,ZH-97} and references therein.}

\begin{figure}[tb]
\centerline{\psfig{width=15truecm,angle=0,file=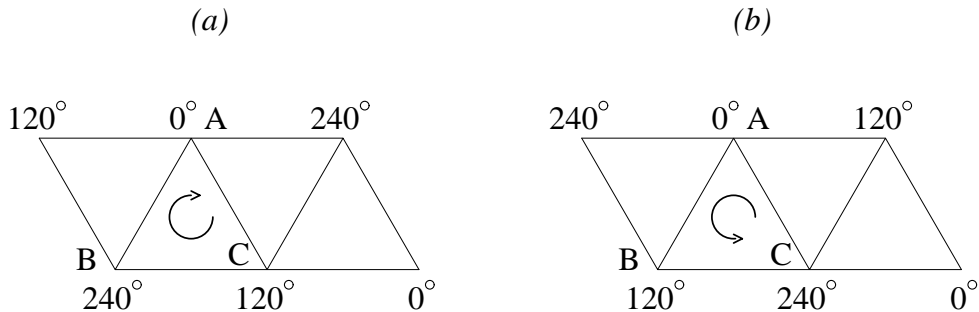}}
\caption{ 
Two inequivalent ground states related by a chiral transformation.
They are obtained as follows: one first fixes $\theta_A = 0^o$,
breaking rotational invariance. Then, there are two possible choices:
on the left we choose $\theta_B = 240^o$, $\theta_C = 120^o$; 
on the right we make the opposite choice. All other lattice spins are 
univocally defined.
}
\label{chiral-tr}
\end{figure}

The Ising model can be generalized by considering the $N$-vector
model on a triangular lattice. In this case one considers
unit $N$-component vectors $\vec{s}_i$ and the Hamiltonian 
\begin{equation}
{\cal H} = J \sum_{\langle ij\rangle} \vec{s}_i\cdot \vec{s}_j.
\label{HFFXYtr}
\end{equation}
Also this model is frustrated: There is no configuration in
which all neighboring spins are antiparallel. However, at variance with the 
Ising case, here the entropy vanishes at zero temperature. Indeed,
once rotational invariance has been broken by fixing the direction of 
one spin, there is a finite 
number of configurations that are global minima of the Hamiltonian
\cite{Villain-77}. For $N=2$, the only case we will consider,
if $\vec{s}_i = (\cos \theta_i,\sin\theta_i)$, one must have
$|\theta_i - \theta_j| = 2 \pi/3$ or $4\pi/3$ when $i$ and $j$ are 
nearest-neighbor sites. It is easy to verify that the degeneracy of the 
ground state is ${\mathbb Z}_2\otimes O(2)$, where $O(2)$ is 
the invariance rotation group. The group ${\mathbb Z}_2$ is due to
the possibility of two (chirally) different configurations.
As shown in Fig.~\ref{chiral-tr}, the ground state is uniquely
determined once one breaks rotational invariance (by setting, for instance,
$\theta_A = 0$) and chooses the chirality of triangle ABC (by setting
$\theta_B = 120^{\rm o}$  or $240^{\rm o}$). An observable that distinguishes 
between the two ground states is the {\it chirality}. Given a 
lattice triangle, see Fig.~\ref{chiral-tr}, we can consider 
\cite{Villain-77}
\begin{equation}
C_n \equiv {2\over 3 \sqrt{3}} [
   \sin(\theta_A - \theta_B) +
   \sin(\theta_B - \theta_C) +
   \sin(\theta_C - \theta_A) ],
\end{equation}
which assumes the values $\pm 1$ on any lattice triangle in a ground-state 
configuration. A good order parameter is obtained as follows. We assign
$s_n = \pm 1$ to each lattice triangle so that $s_n = - s_m$ if triangles
$n$ and $m$ share a lattice link. The order parameter, the {\it chiral
magnetization}, is simply
\begin{equation}
M_C \equiv \sum_n s_n C_n,
\end{equation}
where the sum is extended over all lattice triangles.

\begin{figure}[tb]
\includegraphics[width=17pc]{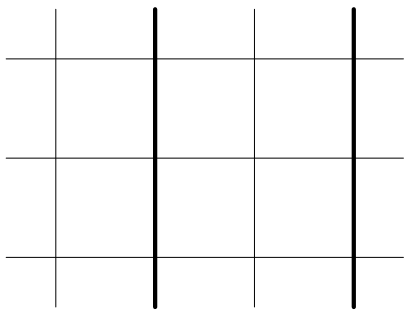}\hspace{2pc}%
\begin{minipage}[b]{17pc}
\caption{\label{fig:FFXY}
The couplings $j_{ij}$ in Hamiltonian~(\protect\ref{FFXY}):
$j_{ij} = 1$ on thin lines, $j_{ij} = -\alpha$ on thick lines.
}
\vspace{10mm}
\end{minipage}
\end{figure}

\begin{figure}[tb]
\centerline{\psfig{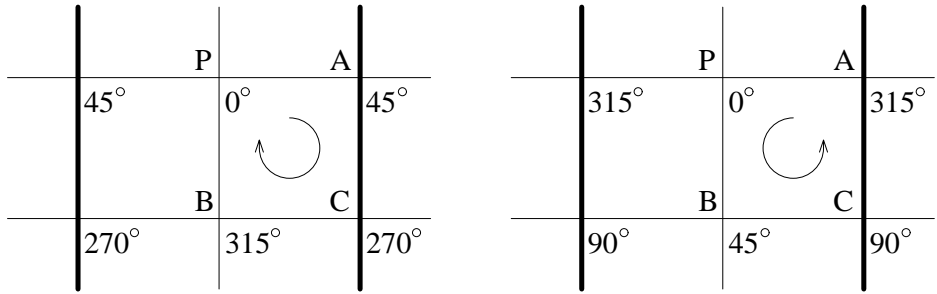}}
\caption{ 
Ground states of the square-lattice FFXY model: in this case 
nearest-neighbor spins must satisfy 
$\theta_i - \theta_j = 45^{\rm o}$ or $315^{\rm o}$ 
if they are connected by a ferromagnetic link,
and $\theta_i - \theta_j = 135^{\rm o}$ or $225^{\rm o}$ 
if they are connected by an antiferromagnetic link.
Once we fix $\theta_P = 0^{\rm o}$, there are two (chirally) 
inequivalent possibilities:
(left) $\theta_A = 45^{\rm o}$ or (right) $\theta_B = 45^{\rm o}$. 
All other spins are fixed. Note that the ground-state configurations
are invariant under translations of two lattice spacings.
}
\label{chiral-sq}
\end{figure}

It is possible to define a frustrated model also on the square lattice.
The relevant model is not the antiferromagnetic Ising or $N$-vector model,
since no frustration occurs on the square lattice or, in general, on any
bipartite lattice. 
To obtain a frustrated model we consider a Hamiltonian of the form
\cite{Villain-77}
\begin{equation}
{\cal H}_{\rm FFXY} = - J 
\sum_{\langle ij\rangle} j_{ij} \, \vec{s}_i \cdot \vec{s}_j ,
\label{FFXY}
\end{equation}
where the two-component spins $\vec{s}_i$ satisfy $\vec{s}_i\cdot \vec{s}_i=1$,
$j_{ij}=1$ along all horizontal lines, while along vertical lines 
ferromagnetic $j_{ij}=1$ and antiferromagnetic $j_{ij}=-\alpha$ 
($\alpha > 0$) couplings alternate, see Fig.~\ref{fig:FFXY}. 
This model is frustrated for any positive $\alpha$. Maximal frustration is 
obtained by taking $\alpha = 1$; for this reason this particular
model, the only one we shall consider in the following,
is called fully frustrated XY (FFXY) model. 
The square-lattice FFXY model 
admits two chirally different ground states, see Fig.~\ref{chiral-sq},
and thus it has the same ground-state degeneracy of the 
antiferromagnetic model on the triangular lattice. These two models
are particular examples of a general class of systems
that all have a ${\mathbb Z}_2\otimes O(2)$ ground-state degeneracy.
We will collectively call them FFXY systems.

Even though the symmetry of the FFXY systems (\ref{HFFXYtr}) and (\ref{FFXY})
is the same as that of the ferromagnetic XY model, we expect the 
critical behavior to be different. Indeed, the universality class
is not only determined by the symmetry of the order parameter but also by the 
symmetry breaking pattern that is different in the two cases.

In order to determine the critical behavior we can perform a direct numerical
study of the model. There is however another possibility, which is the basis
of the field-theoretical approach to critical phenomena. In this case
one first identifies the critical modes of the microscopic Hamiltonian and 
then writes down an effective coarse-grained (continuum) 
Hamiltonian for them. The model 
one obtains is no longer frustrated; still, it is expected to have 
the same critical behavior as the original one. 
In order to derive the effective theory, let us consider the 
antiferromagnetic model on a triangular lattice. As is evident from
Fig.~\ref{chiral-tr}, in the ground state spins rotate by $120^{\rm o}$ 
when moving in the $x$ direction from one site to its neighbor. 
Thus, critical modes are associated with 
fluctuations close to the complex Fourier component $\vec{s}(Q)$ with 
$Q = [2\pi/(3 a),0]$, where $a$ is the lattice spacing. These fluctuations
are parametrized by a complex vector 
$\vec{\Phi} = \vec{\phi}_1 + i \vec{\phi}_2$. Note that the appearance 
of two real two-component fields is at variance with the ferromagnetic case.
Indeed, in that case, the relevant modes are associated with 
the zero-momentum component $\vec{s}(q=0)$. 
As a consequence of the reality condition $\vec{s}(q) = \vec{s}^*(-q)$,
fluctuations are real and are parametrized by a single real 
two-component field.
A standard calculation \cite{CD-85,YD-85,LGK-91}
gives the effective Hamiltonian for the fields
$\vec{\phi}_a$:
\begin{equation}
{\cal H}_{\rm LGW}  = \int d^d x
 \Bigl\{ {1\over2}
      \sum_{a=1,2} \Bigl[ (\partial_\mu \phi_{a})^2 + r \phi_{a}^2 \Bigr]
+ {1\over 4!}u_0 \Bigl( \sum_{a=1,2} \phi_a^2\Bigr)^2
+ {1\over 4} v_0 \phi_1^2 \phi_2^2 \Bigr\}.
\label{HLGW}
\end{equation}
A similar argument applies to the square-lattice FFXY model and gives 
the same effective Hamiltonian (\ref{HLGW}).

Hamiltonian (\ref{HLGW})  has a larger symmetry than the 
original one.
Indeed, the symmetry group is 
$[O(2) \oplus O(2)]\otimes {\mathbb Z}_2$: the $O(2)$ groups are 
related to independent rotations of the two fields, while the 
${\mathbb Z}_2$ group corresponds to the field-interchange symmetry.
Nonetheless---and this is the only property that matters---Hamiltonian
(\ref{HLGW}) for $v_0 > 0$ and FFXY models have the same symmetry-breaking 
pattern. Indeed, for $v_0 > 0$, the ground state corresponds to 
$\phi_1^2 = 0$ and $\phi_2^2\not=0$ or the opposite and thus it has the same 
ground-state degeneracy: the ground state breaks one of the two $O(2)$
groups and the ${\mathbb Z}_2$ field-interchange symmetry.

A lattice discretization of (\ref{HLGW}) is \cite{HPV-05let} 
\begin{eqnarray}
{\cal H}_{\phi} = - 
J \sum_{\langle ij\rangle,a} \vec{\phi}_{a,i}\cdot \vec{\phi}_{a,j} 
+ \sum_{a,i} \left[ \phi_{a,i}^2 + U (\phi_{a,i}^2-1)^2 \right]  + 
2 (U+D) \sum_i \phi_{1,i}^2 \phi_{2,i}^2,
\label{HLGWlat} 
\end{eqnarray}
where $J > 0$ (the model is ferromagnetic),
$a=1,2$, $\vec{\phi}_{a,i}$ is a real two-component variable, the first sum 
goes over all nearest-neighbor pairs, and $\phi^2_a\equiv \vec{\phi}_a \cdot
\vec{\phi}_a$. The correct symmetry-breaking pattern is 
obtained for $D > 0$, which corresponds to $v_0 > 0$ in (\ref{HLGW}).
Moreover, stability requires $U > 0$. For $U\to \infty$,
${\cal H}_\phi$ becomes simpler and we obtain
\begin{equation}
{\cal H} = 
- J \sum_{\langle ij\rangle,a} \vec{\phi}_{a,i}\cdot \vec{\phi}_{a,j} 
+ 2 D \sum_i \phi_{1,i}^2 \phi_{2,i}^2,
\end{equation}
where the fields satisfy the constraint $\phi^2_{1,i} + \phi^2_{2,i} = 1$.
This is the 4-vector model with a spin-4 perturbation that breaks the 
$O(4)$ symmetry to $[O(2)\oplus O(2)]\otimes {\mathbb Z}_2$.
If we additionally take $D\to \infty$ we must have 
$\phi_{1,i}^2 \phi_{2,i}^2 = 0$. In this case, we can parametrize
\begin{equation}
\vec{\phi}_{1,i} = \case{1}{2} (1 + \sigma_i) \vec{s}_i, \qquad\qquad
\vec{\phi}_{2,i} = \case{1}{2} (1 - \sigma_i) \vec{s}_i, \qquad\qquad
\label{mapping}
\end{equation}
where $\sigma_i$ is an Ising spin and $\vec{s}_i$ is a unit two-component
vector. The Hamiltonian reduces to 
\begin{equation}
{\cal H} = 
- \frac{J}{2} \sum_{\langle ij\rangle} 
(1+\sigma_i \sigma_j) \, \vec{s}_i \cdot \vec{s}_j .
\label{IsXY}
\end{equation}
Hamiltonian (\ref{IsXY}) has the same invariance as (\ref{HLGWlat})
although, in terms of the new fields, the $O(2)\oplus O(2)$ symmetry is 
nonlinearly realized:
\begin{eqnarray}
{\vec{s}_i}\! ' &=& [\case{1}{2} (1 + \sigma_i) R_1 + 
              \case{1}{2} (1 - \sigma_i) R_2 ] \vec{s}_i
\\
\sigma_i' &=& \sigma_i,
\end{eqnarray}
where $R_1$ and $R_2$ are $O(2)$ rotation matrices. It is possible to add 
terms of the form $\sigma_i\sigma_j$ without breaking the symmetry 
of the Hamiltonian. We can thus consider the more general Hamiltonian
\cite{GKLN-91}
\begin{equation}
{\cal H}_{\rm IsXY} = 
- \sum_{\langle ij\rangle} 
\left[ \frac{J}{2} (1+\sigma_i \sigma_j) \, \vec{s}_i \cdot \vec{s}_j 
+ C \sigma_i \sigma_j \right] ,
\label{HIsXY}
\end{equation}
We will call this model the Ising-XY (IsXY) model. For $J > 0$ and 
$C + {J}/{2} > 0$ it has the same symmetry-breaking pattern  
as FFXY systems. Thus, it represents another example of this class of models.
We should mention that there are many other systems that share
the same symmetry-breaking pattern: for an extensive list of references
see \cite{HPV-05lungo}.

\section{Results} \label{sec2}

Two-dimensional 
FFXY systems (but note that these systems have also been studied,
both theoretically and experimentally, in three
dimensions \cite{Kawamura-98,PV-review,francesi-review,CPV-04})
have been extensively studied in the last thirty years,
after the appearance of the seminal papers by Villain \cite{Villain-77}.
For an extensive list of references, see \cite{HPV-05lungo}. In spite of that,
their critical behavior is object of debate still today.
Two scenarios have been proposed for the critical behavior of models
(\ref{HFFXYtr}) and (\ref{FFXY}).

A first possibility is that these models have two continuous transitions.
As temperature decreases, there is first a transition associated with
the chiral degrees of freedom: at the transition there is no magnetic 
ordering but only chiral order. As temperature further decreases, there is 
an intermediate phase in which spins are disordered while chiral variables
are magnetized. Then, a second transition occurs, followed by a low-temperature
(LT) phase in which spin-spin correlations decay algebraically.
In this scenario chiral and spin modes do not interact at the transitions
and thus, if the transitions are continuous, one expects a
chiral Ising transition (the order parameter is a scalar) and 
a spin Kosterlitz-Thouless (KT) transition. This scenario has not 
been confirmed numerically so far. The computed exponents at the chiral 
transition do not agree with the Ising ones. For instance, one 
finds $\nu\approx 0.8$ instead of the Ising value $\nu = 1$. 
Moreover, it is not clear how much one can believe in the presence 
of two transitions that are very close to each other: numerically 
$(T_{\rm spin} - T_{\rm chiral}) \simeq 10^{-2} J$.  

The inconsistencies of the two-transition scenario apparently favor 
the presence of a single critical point. In this scenario, the observed 
difference between the critical temperatures is interpreted as a 
correction-to-scaling effect. Since chiral and spin modes become critical
at the same temperature, it is possible that the transition belongs to 
a new universality class with a new set of critical exponents. 
In this scenario, the result $\nu\approx 0.8$ would be fully
acceptable. Of course, it is also possible to interpret the 
results for $\nu$ in terms of crossover effects. Since the two transitions
are very close, scaling corrections may be large so that the asymptotic 
behavior can be observed only on very large lattices \cite{Olsson-95}.

We have considered again the issue \cite{HPV-05let,HPV-05lungo},
performing extensive simulations of the
FFXY model (\ref{FFXY}), of the $\phi^4$ model (\ref{HLGWlat}) with 
$U=1$ and of the IsXY model (\ref{HIsXY}). In all cases 
we have considered the square lattice. We have used a mixture
of Metropolis and overrelaxation updates as well as cluster updates in the 
LT phase, essentially following \cite{GPP-98} (see \cite{HPV-05lungo} for a 
detailed discussion).  We have studied the 
critical behavior on lattices of size $L\times L$, in some cases
up to $L\sim 10^3$: for the FFXY model, 
the $\phi^4$ model with $D = 1/2$, and the IsXY model with $C = 0$, the 
largest lattice we have used at the chiral transition corresponds to $L=1000$,
$L=1200$, $L=360$ respectively. In the LT phase the 
Monte Carlo algorithm is much more efficient and we have been able
to simulate even larger sizes: 
for the $\phi^4$ model with $D=1/2$ we performed simulations for $L=2048$. 

The analysis of the Monte Carlo 
results for the square-lattice FFXY model definitely shows that this model 
undergoes two transitions: the chiral one belongs to the Ising universality 
class, while the spin one is compatible with a KT behavior. 
In the LT phase the critical
behavior of the spin modes is controlled by the same line of Gaussian 
fixed points as in the standard XY model. The discrepancies from Ising behavior 
at the chiral transition that have been observed in previous 
studies are simply crossover effects. They are due to the presence of 
a large, albeit finite spin correlation length $\xi_s^{(c)}$ at the 
chiral transition. In finite-size scaling studies the 
asymptotic behavior can only be observed if $L\gg \xi_s^{(c)}$.
Since $\xi_s^{(c)}$ is quite large, $\xi_s^{(c)} = 118(1)$, the 
asymptotic behavior can only be observed in simulations 
with $L\simeq 500$-1000, i.e. for values of $L$ that are 
much larger than those that could be used in simulations until 
a few years ago. In the other models we have studied, 
the determination of the asymptotic behavior may be even more difficult.
For instance, in the $\phi^4$ model with $D=1/2$, $\xi_s^{(c)} \simeq 380$. 
Thus, even with simulations with $L=1200$, we have not been able to 
observe the Ising behavior but only the beginning of the crossover 
towards the asymptotic behavior.

\begin{figure}[tb]
\begin{minipage}{17pc}
\includegraphics[width=17pc]{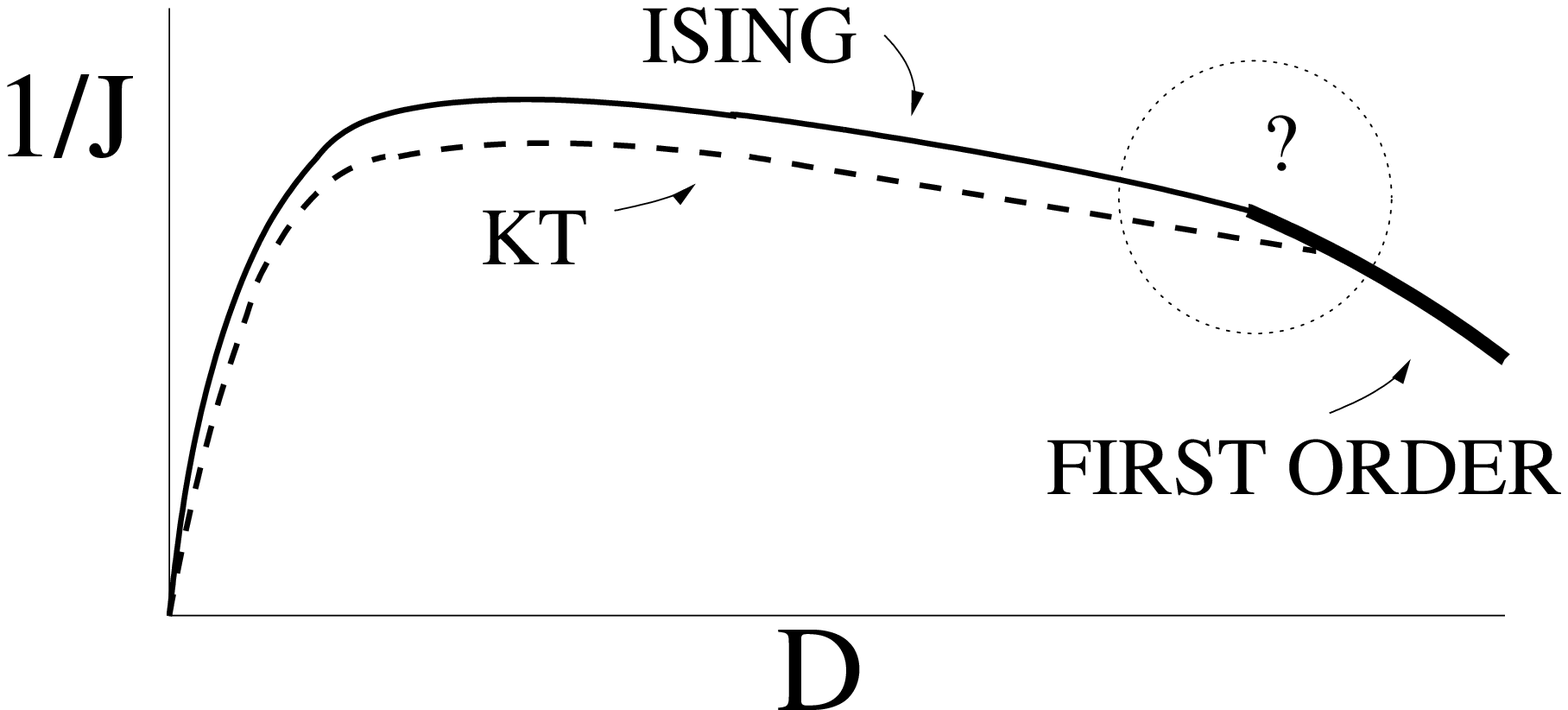}
\end{minipage}\hspace{2pc}%
\begin{minipage}{17pc}
\includegraphics[width=17pc]{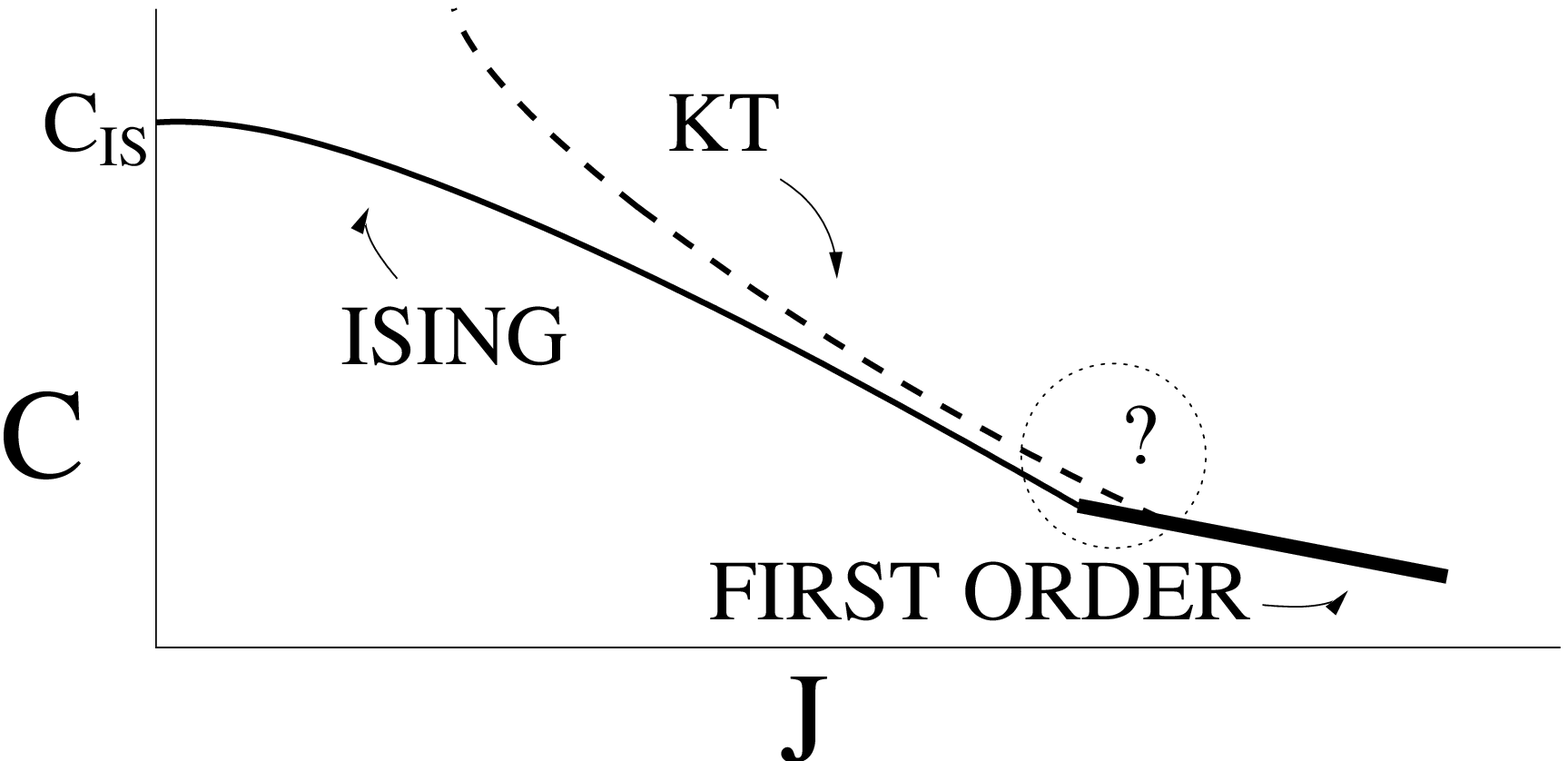}
\end{minipage} 
\caption{\label{phasediag}
Sketch of the phase diagram of the $\phi^4$ model (\protect\ref{HLGWlat}) 
for $U=1$ and $D > 0$ (left) and of the IsXY model 
(\protect\ref{HIsXY}) (right). 
The continuous, dashed, and thick continuous 
lines represent  Ising, KT, and first-order transition lines.
The distance between the ferromagnetic Ising and KT lines is amplified;
otherwise, the two transitions cannot be distinguished on the scale of the 
figure.  The phase diagram within the circled region is unknown. 
In the IsXY case there is also an antiferromagnetic Ising transition (af)
starting at $C = - C_{\rm Is} = - {1\over2}(1 + \sqrt{2})$, $J=0$.
}
\end{figure}

In the IsXY and in the $\phi^4$ model we have observed two transitions
in a large parameter region. However, for $D$ or $-C$ large,
we have also observed a unique first-order transition which separates
the LT phase with chiral order and spin quasi-long-range order from the 
disordered phase, see Fig.~\ref{phasediag}. Thus, our results confirm 
the two-transition scenario for generic FFXY systems 
in the sense that we have found no 
evidence of a unique {\it continuous} transition where chiral and 
spin modes become both critical. A single transition occurs
only if it is of first order.

Beside these results that confirm the two-transition scenario, we have also
observed an unexpected universal crossover behavior.
We find that renormalization-group invariant quantities
(e.g., critical exponents, Binder parameters, $\ldots$) 
computed in the different models scale 
at the chiral and spin transitions respectively as 
\begin{equation}
{\cal R} = f_{\cal R}^{(c)} (L/l), \qquad\qquad
{\cal R} = f_{\cal R}^{(s)} (L/l),
\label{scaling}
\end{equation}
where $l$ is a model-dependent scaling factor that is identical at the two
transitions. At the chiral transition---the only case in which
we have done a systematic investigation by varying $C$ and $D$---corrections
appear to increase as $D$ or $-C$ increases 
(in practice, significant deviations are observed for $D \gtrsim 4$ 
and $C \lesssim -2$). 
Of course, it is of interest to have 
a renormalization-group explanation of this apparent universality.
Eq.~(\ref{scaling}) can be explained by the presence of a multicritical 
point \cite{Amit-book,LS-84,KNF-76} 
or a line of multicritical points of the same type) where 
chiral and spin modes become both critical. Indeed, close to 
a multicritical point we expect that any RG-invariant quantity behaves as
\begin{equation}
{\cal R} = \hat{f}_{\cal R} (L/\xi_s, L/\xi_{\rm ch}),
\label{MCR-general}
\end{equation}
where $\xi_s$ and $\xi_{\rm ch}$ are the infinite-volume 
correlation lengths for spin and chiral variables.
Our analysis at the Ising and KT transitions corresponds to fixing 
$L/\xi_{\rm ch} = 0$ and $L/\xi_{s} = 0$, i.e. provides the scaling function 
along two particular lines. 
If the interpretation in terms of a multicritical point
is correct, the functions $f_{\cal R}(x)$  provide informations on 
the behavior at the multicritical point. Indeed, while Ising or KT behavior is 
observed for $x\to\infty$, in the opposite limit $x\to 0$ we 
obtain the value of the RG-invariant quantity  $\cal R$ 
at the multicritical point.

The nature of the
multicritical point is unclear. One possibility 
is the $O(4)$ multicritical point that is present in the 
$\phi^4$ theory for $D = 0$. Another possibility is the 
multicritical point that appears in frustrated XY systems with 
modulated couplings (for instance in model (\ref{FFXY}) for $\alpha \not=1$)
\cite{BDGL-86,EHKT-89,GKS-98} or in generalizations
of the IsXY model, in which an additional spin-spin coupling is 
added, breaking the $O(2)\oplus O(2)$ symmetry.

Finally, we wish to compare with field-theory (FT) approaches. 
Perturbative analyses \cite{CP-01} 
of model (\ref{HLGW}) indicate the existence of 
a new universality class associated with the symmetry-breaking pattern
$[O(2)\oplus O(2)]\otimes {\mathbb Z}_2\to O(2)$.  Even though we have found
no evidence for it, our results do not necessarily contradict those 
of Ref.~\cite{CP-01}. It is possible that the models we have considered 
are outside the attraction domain of the FT fixed point. If this is the case,
field theory provides another candidate for the multicritical point. 
The models we consider could be outside, but close to the attraction domain
of the fixed point---this is not unplausible since $\xi_s^{(c)}$ is 
large---so that the crossover behavior is controlled by the FT fixed point.
We wish also to make a remark on the validity of (\ref{HLGW}).
In the derivation of Hamiltonian (\ref{HLGW})
by using the standard Hubbard-Stratonovitch transformation, 
terms with more than four fields are neglected \cite{CD-85,YD-85}.
In particular, terms of the form $ (\vec{\phi}_1 \cdot \vec{\phi}_2)^n$ 
appear at sixth order $(n=3)$ (resp. eighth order) 
in the case of the triangular-lattice (resp. square-lattice) FFXY model
\cite{YD-85}. These terms have only ${\mathbb Z}_2 \oplus O(2)$ symmetry
and thus, under renormalization-group transformations, are bound to 
generate a term of the form $(\vec{\phi}_1 \cdot \vec{\phi}_2)^2$, or 
even a quadratic term
of the form $(\vec{\phi}_1 \cdot \vec{\phi}_2)$. We obtain therefore the
multicritical Hamiltonian
\begin{equation}
{\cal H}_{\rm LGW,2}  = 
{\cal H}_{\rm LGW} + \int d^d x 
  \left[
      {1\over2} r_2 (\vec{\phi}_1 \cdot \vec{\phi}_2) 
    + {1\over 4} z_0 (\vec{\phi}_1 \cdot \vec{\phi}_2)^2
    + {1\over 4} z_1 (\vec{\phi}_1 \cdot \vec{\phi}_2) (\phi_1^2 + \phi_2^2) 
     \right],
\label{HLGW2}
\end{equation}
as it has been postulated for modulated systems.
If this interpretation is correct, the FT fixed point may only be relevant
at the multicritical point, provided it is stable under the quartic 
perturbations that break the 
$O(2)\oplus O(2)$ symmetry. This holds in three dimensions \cite{PV-04}, 
but nothing is known in the two-dimensional case. It should be remarked 
that these considerations are only relevant for the FFXY model. 
The IsXY and $\phi^4$ theories {\em are} $O(2)\oplus O(2)$ invariant and thus
the correct FT Hamiltonian is clearly (\ref{HLGW}) and not (\ref{HLGW2}).

\section{Chiral transition} \label{sec3}

In order to determine the nature of the chiral transition,
we have studied the behavior of several quantities 
at fixed $R_c\equiv \xi_c/L$ where $\xi_c$ is the chiral correlation length
(see \cite{HPV-05lungo} for a precise definition in the different models).
We use the method proposed in  \cite{Hasenbusch-99} and 
further discussed in \cite{CHPRV-01}. 
We fix $R_c$ equal to $R_{\rm Is}$, where $R_{\rm
Is}=0.9050488292(4)$ is the universal value of $\xi/L$ at the critical
point in the 2-$d$ Ising universality class \cite{SS-00}. 
We stress that this choice does not bias our analysis in favor of the Ising
nature of the chiral transition. For any chosen value 
(as long as it is positive) and whatever the 
universality class of the chiral transition is (it may also coincide 
with the spin transition), we are studying the 
model for $L$-dependent temperatures $T_{\rm eff}(L)$ such that 
$T_{\rm eff}(L) \to T_{\rm ch}$ for $L\to \infty$. 
Note, however, that quantities like the Binder parameter depend 
on the chosen value for $R_c$. Indeed, in the finite-size scaling limit,
$R_c = f_R[L^{1/\nu} (T - T_{\rm ch})]$. Therefore, fixing $R_c$ is 
equivalent to fixing $X \equiv L^{1/\nu} (T_{\rm eff}(L) - T_{\rm ch})$.
Since the Binder parameter satisfies an analogous relation 
$B_c = f_B[L^{1/\nu} (T - T_{\rm ch})]$, at fixed $R_c$ 
$B_c$ converges to $f_B(X)$. By fixing $R_c$ to the critical Ising value,
we will be able to perform an additional consistency check.
If the chiral transition belongs to the Ising universality 
class, then $X = 0$ (apart from scaling corrections) and we should find that 
any RG-invariant quantity 
converges to its critical-point value in the Ising model.

\begin{figure}[tb]
\centerline{\psfig{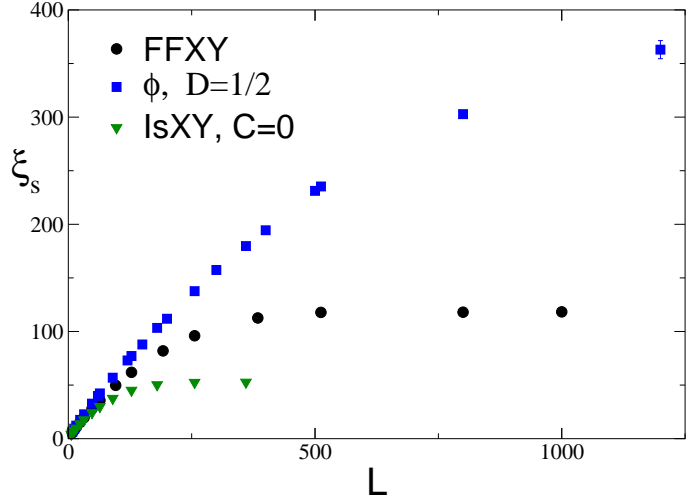}}
\caption{ 
Spin correlation length $\xi_s$ at fixed $R_c=R_{\rm Is}$ 
(chiral transition). 
For $L\to \infty$, we have $\xi_s = 118(1)$ in the FFXY model,
$\xi_s = 52.7(4)$ in the IsXY with $C=0$, and 
$\xi_s \approx 380$ in the $\phi^4$ model with $U=1$ and $D=1/2$. 
}
\label{xis-chiral}
\end{figure}

\begin{figure}[tb]
\centerline{\psfig{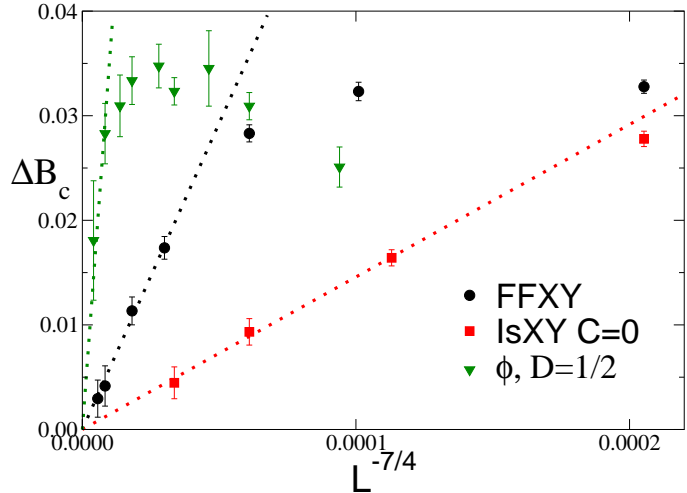}}
\vspace{1mm}
\caption{Chiral Binder parameter $B_c$ at fixed
$R_c=R_{\rm Is}$ (chiral transition).  Plot of 
$\Delta B_c\equiv B_c - B_{\rm Is}$ at fixed $R_c=R_{\rm Is}$ vs $L^{-7/4}$,
for the FFXY model, the $\phi^4$ model at $D=1/2$, and the IsXY model
at $C=0$. $B_{\rm Is} = 1.167923(5)$ is the value of the 
Binder parameter at the critical point in the Ising model \protect\cite{SS-00}.
}
\label{bc-chiral}
\end{figure}

We first verify that $\xi_s$ converges to a constant as $L\to\infty$.
In Fig.~\ref{xis-chiral} we show the numerical results. The correlation length
is clearly finite in the FFXY model and in the IsXY model with $C=0$.
In the $\phi^4$ model with $D=1/2$ we do not yet observe that 
$\xi_s$ is finite, although it is already clear that $\xi_s$ does not 
increase linearly with $L$, as it would be the case if 
the spin correlation length were infinite for $L=\infty$.
Then, we verified that the transition, if continuous, belongs to the Ising 
universality class. The best evidence is provided by the Binder chiral
parameter. If the transition belongs to the Ising universality 
class we should find \cite{SS-00,CHPV-02,HPV-05lungo}
\begin{equation}
B_c = B_{\rm Is} + b L^{-7/4},
\end{equation}
where $B_{\rm Is} = 1.167823(5)$ \cite{SS-00}. The results reported in
Fig.~\ref{bc-chiral} are fully consistent with Ising behavior for 
$L\gg \xi_s^{(c)}$. Not only do we observe the Ising asymptotic value,
but also the rate of convergence is well verified. Also the 
critical exponents $\nu$ and $\eta$ converge to the Ising values
for $L\gg \xi_s^{(c)}$ (results for $\nu$ will be shown below).

\section{Low-temperature phase and spin transition} \label{sec4}

Since the spin correlation length is finite at the chiral transition there 
must be a paramagnetic phase with chiral order. Such a phase ends at a 
second transition which is followed by the LT phase in which
chiral order and spin quasi-long-range order coexist.

We first study the nature of the LT phase and
verify the breaking of the ${\mathbb Z}_2$ invariance. 
Direct evidence is provided by the chiral Binder parameter $B_c$.
If chiral modes are magnetized, $B_c\to 1 + O(L^{-2})$ for
$L\to \infty$. This behavior is very well verified in all models.
In the $\phi^4$ model, the ${\mathbb Z}_2$ group corresponds to the 
field-interchange symmetry. Thus, the ${\mathbb Z}_2$ symmetry breaking 
implies that only one of the two fields $\phi_1$ and $\phi_2$ 
is critical in the LT phase. Our numerical results fully confirm this
expectation. One can distinguish the fields according to the 
value of $Q_a = \sum_i \phi^2_a$. The field with the largest value of $Q$
is critical (for instance, the corresponding susceptibility 
and correlation length diverge as $L\to \infty$),
while the other one is not (the susceptibility and the correlation length 
have a finite limit as $L\to \infty$).

\begin{figure}[tb]
\centerline{\psfig{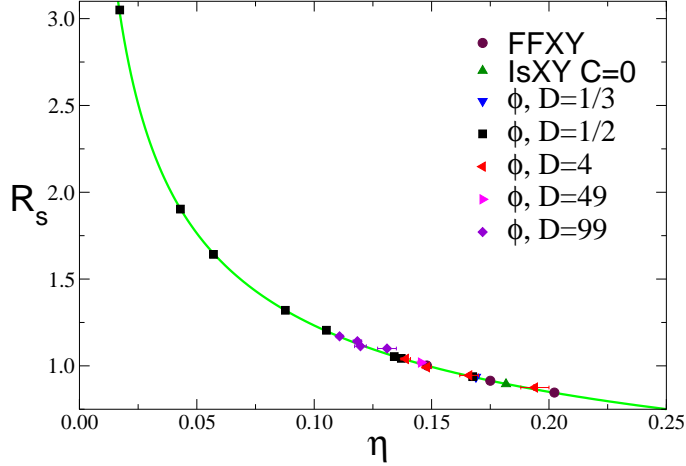}}
\vspace{2mm}
\caption{ 
Estimates of $R_s\equiv \xi_s/L$ vs $\eta$ in the LT phase. 
The continuous line is the prediction obtained by assuming that in the 
LT phase criticality is controlled by the same line of Gaussian fixed points
as in the XY model.}
\label{xil-vs-eta}
\end{figure}

Once we have checked that chiral modes are magnetized (this is of course 
obvious because of the presence of the chiral transition), we 
study the behavior of the spin variables and 
verify that the large-$L$ behavior is controlled by the same line of 
Gaussian fixed points as in the 
standard XY model. In the XY model one can derive universal relations
among renormalization-group invariant quantities that are valid
in the whole LT phase, up to the KT transition \cite{Hasenbusch-05}.
Indeed, below the KT transition the spin-wave approximation is asymptotically 
exact as $L\to\infty$ and allows an analytic determination of any quantity
in terms of the spin-wave parameter. Such a parameter is not universal
and can be eliminated by expressing a renormalization-group invariant
quantity in terms of another. For instance, one can express
the helicity modulus on a square lattice $L\times L$ with periodic
boundary conditions in terms of the exponent $\eta$ computed from
the size dependence of the magnetic susceptibility ($\chi\sim L^{2-\eta}$):
\begin{equation}
\Upsilon = {1\over 2\pi \eta} - 
    {\sum_{n=-\infty}^\infty n^2 \exp(-\pi n^2/\eta) \over 
    \eta^2 \sum_{n=-\infty}^\infty \exp(-\pi n^2/\eta) },
\end{equation}
where $0< \eta \le 1/4$.
Analogously, one can express $\xi_s/L$ in terms of $\eta$.
In Fig.~\ref{xil-vs-eta} we compare the spin-wave prediction with
numerical results. The agreement is quite good, confirming that 
FFXY systems and the standard XY model have the same
LT phase, as far as the spin degrees of freedom are concerned. 

The above-reported results for the LT phase make it plausible that 
the spin transition belongs to the KT univerality class. Another 
check is provided by our numerical results 
for $\xi_s/L$ and $\Upsilon$. In the XY model, at the KT transition
\cite{Hasenbusch-05}
$\xi_s/L \approx 0.750691 + 0.212430/\ln(L/C_1)$ and 
$\Upsilon \approx 0.636508 + 0.318899/\ln(L/C_2)$. 
In all models we have studied 
these two quantities assume the XY values approximately at the 
same temperature---which we identify with the spin
critical temperature---thereby confirming the KT nature of the 
transition.

\section{Crossover behavior} \label{sec5}

As we already mentioned, our data show the scaling behaviors (\ref{scaling}).
In this section we shall give a few details. 

\begin{figure}[tb]
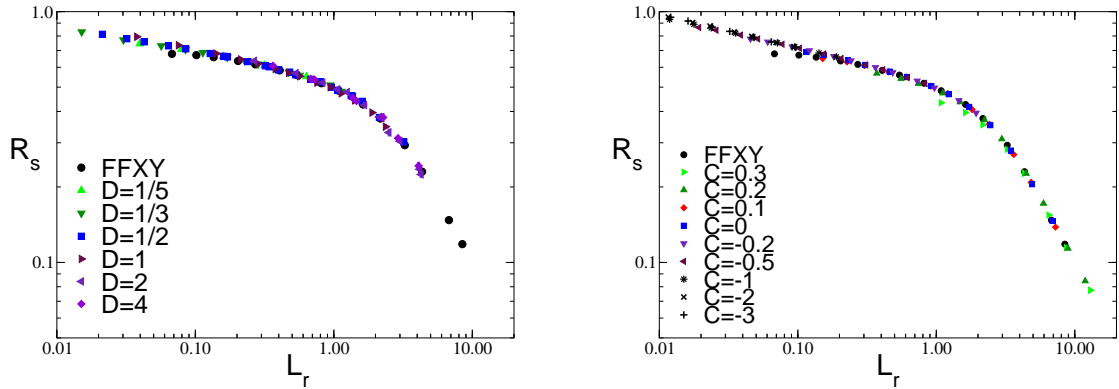

\begin{minipage}{17pc}
\includegraphics[width=16pc]{xillog.eps}
\end{minipage}\hspace{2pc}%
\begin{minipage}{17pc}
\includegraphics[width=16pc]{xilisxylog.eps}
\end{minipage} 
\caption{ 
Ratio $R_s\equiv \xi_s/L$ ($\xi_s$ is the spin correlation 
length) at the chiral transition vs $L_r \equiv L/l$
for the FFXY model and $\phi^4$ (left) and IsXY (right) models. 
We set $l=\xi_s^{(c)}$ for the FFXY model; the values of $l$ for the 
other models are obtained by requiring that all data fall
on a single curve. Note the logarithmic scale on both axes.
For $L_r\to \infty$, $R_s$ converges to 0.
}
\label{xisul-crossover}
\end{figure}

In order to verify the scaling behavior (\ref{scaling}) at the 
chiral transition\footnote{
Note that our results have been obtained at fixed $R_c$ and not at the 
chiral critical point. However, since we are dealing with an Ising
transition and $R_c$ has been fixed to the critical-point Ising 
value, this is irrelevant in the scaling limit. Had we fixed $R_c$ 
to a different value, we would have obtained quantitatively different
scaling curves, corresponding to  
the limit $\xi_{\rm ch},L\to \infty$ at fixed $L/\xi_{\rm ch}\not=0$
in Eq.~(\ref{MCR-general}) ($\xi_{\rm ch}$ is the infinite-volume 
chiral correlation length).
}
we have first considered the 
data for $R_s\equiv \xi_s/L$ ($\xi_s$ is the spin correlation length)
at the chiral transition and we have investigated whether 
they fall on a single curve by using a rescaled 
variable $L_r \equiv L/l$, where $l$ is a rescaling factor that 
depends on the model. The result are reported in Fig.~\ref{xisul-crossover}.
The data fall on a single curve with remarkable precision, i.e. 
$\xi_s/L = f_s(L/l)$ where $f_s(x)$ is model independent. Note that 
the rescaling factors change significantly from one model to another:
for instance, $l/l_{\rm FFXY} = 0.75$ (resp. 7.0) in the $\phi^4$ model
with $D = 4$ (resp. $D = 1/5$) and 
$l/l_{\rm FFXY} = 0.031$ (resp. 12) in the IsXY model
with $C = 0.3$ (resp. $C = -2$). It is easy to realize that $l$ should 
be proportional to $\xi_s^{(c)}$, the infinite-volume spin correlation length
at the chiral transition. Indeed, since $\xi_s \to \xi_s^{(c)}$ as 
$L\to \infty$, we have $f_s(x) \sim a/x$ for $x\to \infty$, where 
$a$ is model independent ($f_s(x)$ is model independent). Moreover,
$\xi_s^{(c)} = l/a$. Therefore, if we fix $l = \xi^{(c)}_s$ in 
one model, then the same holds in all different models. In the figures
we have chosen $l_{\rm FFXY} = 118 \approx \xi_s^{(c)}$ and thus the plots
we present are indeed in terms of $L/\xi_s^{(c)}$, even in those cases
in which we have not been able to determine directly the spin
correlation length for $L\to \infty$.

\begin{figure}[tb]
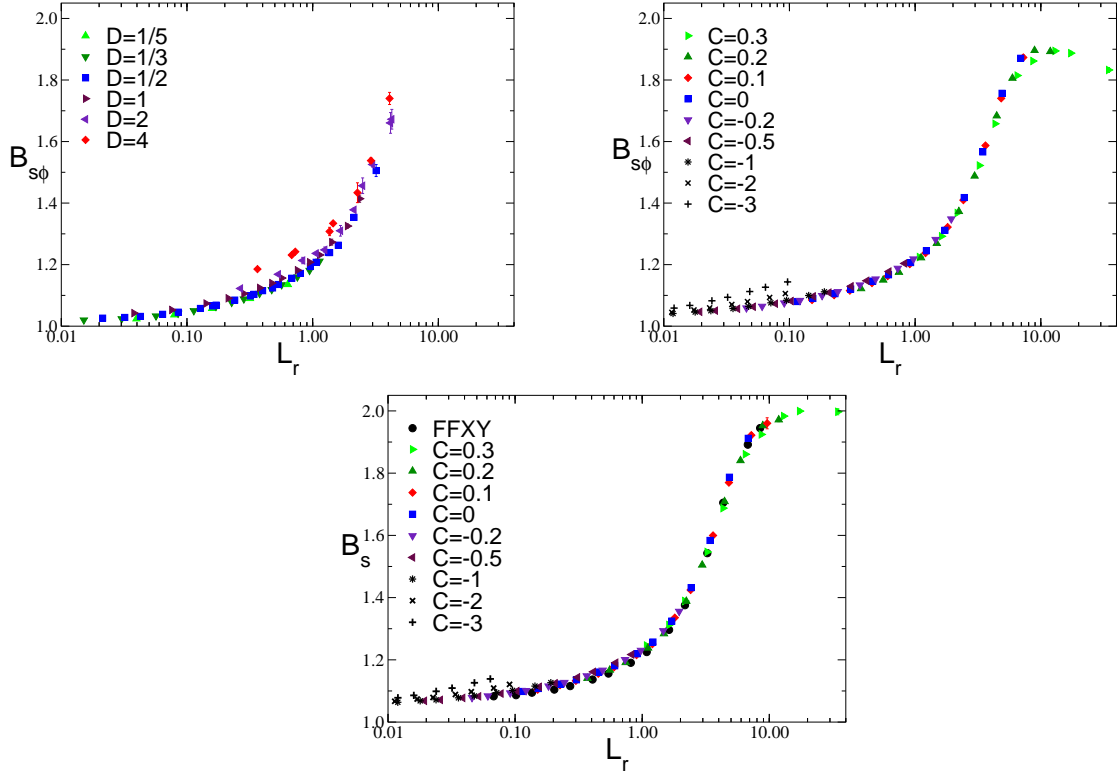

\begin{minipage}{17pc}
\includegraphics[width=16pc]{bi1.eps}
\end{minipage}\hspace{2pc}%
\begin{minipage}{17pc}
\includegraphics[width=16pc]{bi2.eps}
\end{minipage} 
\begin{center}
\begin{minipage}{17pc}
\includegraphics[width=16pc]{bi3.eps}
\end{minipage} 
\end{center}
\caption{ 
Spin Binder parameters $B_s$ and $B_{s\phi}$ 
at the chiral transition vs $L_r\equiv L/l$. Results
for the FFXY, $\phi^4$, and IsXY models. 
For $L_r\to \infty$, $B_s$ and $B_{s\phi}$ converge to 2 and 3/2 
respectively.
The rescalings $l$ are the same as in Fig.~\protect\ref{xisul-crossover}.  
}
\label{bs-crossover}
\end{figure}

\begin{figure}[tb]
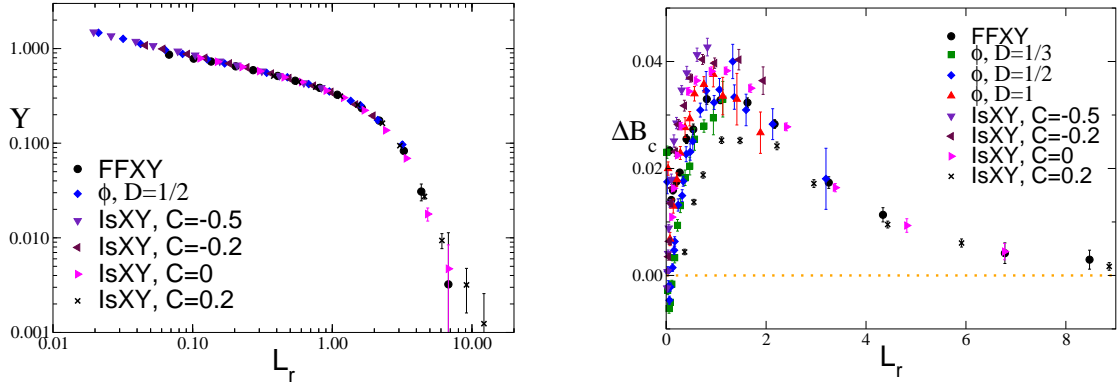

\begin{minipage}{17pc}
\includegraphics[width=16pc]{yupsilon.eps}
\end{minipage}\hspace{2pc}%
\begin{minipage}{17pc}
\includegraphics[width=16pc]{bicr.eps}
\end{minipage}
\caption{
Helicity modulus $\Upsilon$ (left) and 
chiral Binder parameter $B_c$ (right) at the chiral transition 
vs $L_r \equiv L/l$. We report
$\Delta B_c = B_c - B_{\rm Is}$, where
$B_{\rm Is}$ is the value of the Binder parameter at the
critical point in the Ising model.
The rescalings $l$ are the same as in Fig.~\protect\ref{xisul-crossover}.  
}
\label{Upsilon-crossover}
\end{figure}

In order to verify the universality of the scaling Ansatz (\ref{scaling}),
we have considered the spin Binder parameters $B_{s\phi}$ and $B_s$
(we use two different inequivalent definitions, see
\cite{HPV-05lungo}). In Fig.~\ref{bs-crossover}
we plot $B_s$ and $B_{s\phi}$ versus $L/l$. 
We use the rescaling factors that have been
determined in the analysis of $R_s$. The agreement is quite good. 
Deviations appear as $D$ or $-C$ increases. In particular, the data 
for $D = 4$ ($\phi^4$ model) and for $C = -3$ (IsXY model) are outside the 
curve. They would fall on the same curve as the others, only if the 
rescaling factor is changed by a factor of 2 and 5 respectively in the two 
cases. In Fig.~\ref{Upsilon-crossover} we plot the results for the 
helicity modulus: again all data fall on a single curve quite precisely.
It is interesting to note that, for $0.02\lesssim L_r \lesssim 0.5$
$R_s$ and $\Upsilon$ show an approximate power-law behavior: they
behave as $L^{-\epsilon}_r$ with $\epsilon\approx 0.1$ ($R_s$) and 
$\epsilon\approx 0.33$ ($\Upsilon$). If this behavior holds also
for smaller values of $L_r$, we have 
$\Upsilon, R_s \to \infty$ for $L_r \to 0$.

\begin{figure}[tb]
\centerline{\epsfig{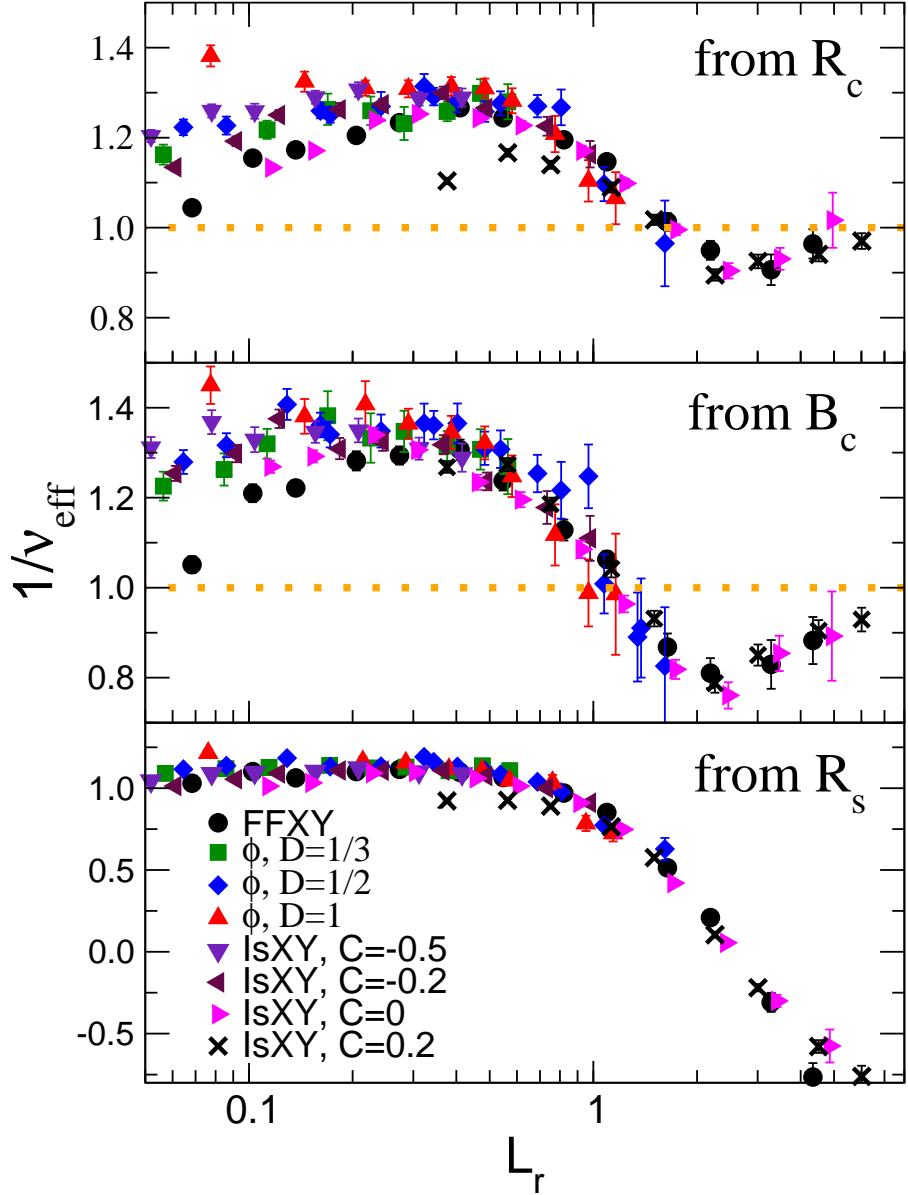}}
\caption{Effective exponent 
$1/\nu_{\rm eff}$ computed by using $R_c\equiv \xi_c/L$, the 
Binder chiral parameter $B_c$, and $R_s\equiv \xi_s/L$ 
($\xi_c$ and $\xi_s$ are respectively the chiral and spin 
correlation lengths) vs
$L_r\equiv L/l$ at the chiral transition. 
For $L_r\to\infty$, $1/\nu_{\rm eff}$ converges to
$1/\nu_{\rm Is} = 1$ for $R_c$ and $B_c$, and $-1$ for $R_s$.  
The rescalings $l$ are the same as in Fig.~\protect\ref{xisul-crossover}.  
}
\label{nueff}
\end{figure}

Next we consider the chiral variables. They also show the universal behavior 
(\ref{scaling}). 
In Fig.~\ref{Upsilon-crossover} we report the Binder chiral parameter,
with the same scaling factors $l$ as before. The data fall again on a single
curve although the curve does not change significantly as $L/l$ varies,
at variance with the spin variables.

Finally, we show the effective exponent $\nu_{\rm eff}$ that can be 
obtained from the derivatives of $R_c$, $B_c$, and $R_s$. 
The exponent obtained from the chiral variables should converge to the 
Ising value $\nu = 1$ as $L\to \infty$, and indeed it does. 
The approach is however 
nonmonotonic, $\nu_{\rm eff}$ being first smaller than 1, then larger. 
It is interesting to note that for $L_r \lesssim 1$, 
i.e.~$L\lesssim \xi_s^{(c)}$, $\nu_{\rm eff}$ is approximately constant 
and equal to 0.8. This behavior explains previous results. Indeed, if 
one performs simulations only for values of $L$ such that 
$L_r \lesssim 1$ (this was the case in previous simulations of the 
FFXY model since 
$\xi_s^{(c)} \approx 10^2$) one would estimate $\nu =0.8$.
Note also that, for $L\lesssim \xi_s^{(c)}$, the effective exponent
$\nu_{\rm eff}$ obtained from the spin variable $R_s$ is approximately
constant and close to the value obtained by using chiral
variables. In this range of values of $L$ chiral and spin variables
appear as if they are both critical.

\begin{figure}[tb]
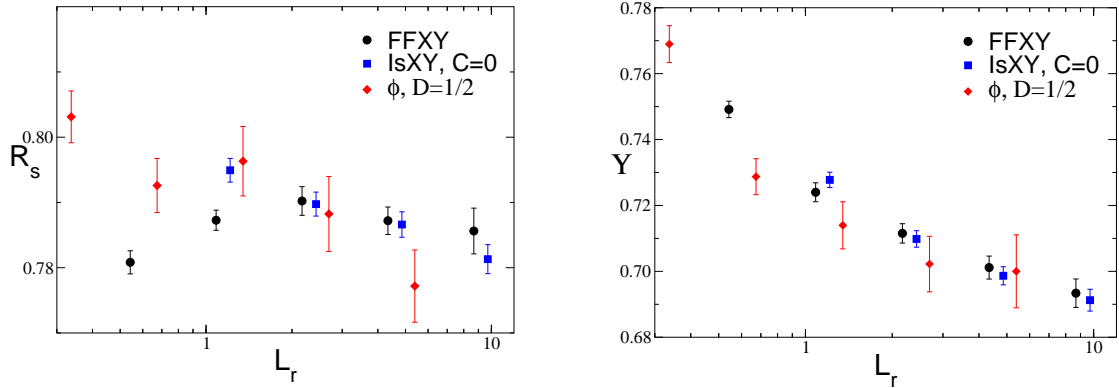

\begin{minipage}{17pc}
\includegraphics[width=16pc]{xisprs2.eps}
\end{minipage}\hspace{2pc}%
\begin{minipage}{17pc}
\includegraphics[width=16pc]{ysprs2.eps}
\end{minipage}
\caption{
Ratio $R_s\equiv \xi_s/L$ (left) and helicity modulus $\Upsilon$ (right) 
at the spin transition vs $L_r \equiv L/l$. 
The rescalings $l$ are the same as in Fig.~\protect\ref{xisul-crossover}.  
}
\label{spin-crossover}
\end{figure}

The same analysis can be repeated at the spin transition. In
Fig.~\ref{spin-crossover} we report $\xi_s/L$ and $\Upsilon$
versus $L/l$ where the rescaling factors $l$ are those determined at the chiral
transition. The agreement is again quite good. The existence of scaling 
at the two transitions with the same rescaling factors is another 
piece of evidence in favor of the multicritical origin of 
the universality we observe.

The crossover curves we have computed can give us some hints on the nature
of the multicritical point, if it really exists. Indeed, the 
behavior at the multicritical point is simply obtained by considering 
the limit $L_r \to 0$. 
First, let us notice that the data at the chiral transition 
apparently exclude the possibility of a decoupled multicritical point
in which spin and chiral modes have XY and Ising behavior.
Indeed, the helicity modulus is 
very much different from the KT value,
$\Upsilon_{KT} = 0.63650817819\ldots$
\cite{Hasenbusch-05} and $B_c$ is apparently smaller than
the Ising value for $L_r \to 0$.
O(4) behavior is possible, since, as we already discussed,
our data are compatible with
$\Upsilon, R_s \to \infty$ for $L_r \to 0$. Also the data for the 
Binder parameters $B_s$ and $B_{s\phi}$ are compatible with the O(4) 
value, $B_s = B_{s\phi} = 1$.  For $L_r\to 0$, the crossover
curves at the spin transition should converge to the same values as 
those at the chiral critical point. This is not evident from the 
results plotted in Fig.~\ref{spin-crossover}.
This can be easily explained. 
At the spin transition the natural scale is the 
infinite-volume chiral correlation length
at the transition $\xi_c^{(s)}$. We expect XY behavior for 
$L/\xi_c^{(s)} \gg 1$ and multicritical behavior in the opposite case. 
Numerically, we find $\xi_s^{(c)}/\xi_c^{(s)} \approx 15$, so that 
$L/\xi_c^{(s)} = 1$ corresponds to $L_r = L/\xi_s^{(c)} \approx 0.07$.
As it can be seen from the figure, none of our data satisfies the 
condition $L_r \ll 0.07$, so that we are unable to observe 
multicritical behavior at the spin transition.


\section*{References}

\begin{thebibliography}{99}

\bibitem{Wannier-50} 
Wannier G H 1950 
Antiferromagnetism. The triangular Ising net
{\it Phys. Rev.} {\bf 79} 357;
erratum 1973 {\it Phys. Rev.} B {\bf 7} 5017

\bibitem{Houtappel-50}
Houtappel R M F 1950 
Order-disorder in hexagonal lattices
{\it Physica} {\bf 16} 425

\bibitem{NMH-93}
Nagai O, Miyashita S and Horiguchi T 1993 
Ground state of the antiferromagnetic Ising model of general spin $S$
on a triangular lattice
{\it Phys. Rev.} B {\bf 47} 202

\bibitem{LHL-95}
Lipowski A, Horiguchi  T and Lipowska D 1995
Critical behavior of spin $S$ antiferromagnetic Ising model on triangular 
lattice
{\it Phys. Rev. Lett.} {\bf 74} 3888

\bibitem{ZH-97}
Zeng C and Henley C L 1997 
Zero-temperature phase transitions of an antiferromagnetic Ising model of 
general spin on a triangular lattice
{\it Phys. Rev.} B {\bf 55} 14935

\bibitem{Villain-77}
Villain J 1977 
Spin glass with non-random interactions
{\it J. Phys. {\rm C}: Solid State Phys.} {\bf 10} 1717; 
Villain J 1977 
Two-level systems in a spin-glass model. I.
General formalism and two-dimensional model
{\it J. Phys. {\rm C}: Solid State Phys.} {\bf 10} 4793

\bibitem{CD-85}
Choi M Y and Doniach S 1985 
Phase transitions in uniformly frustrated XY models
{\it Phys. Rev.} B {\bf 31} 4516

\bibitem{YD-85}
Yosefin M and Domany E 1985 
Phase transitions in fully frustrated spin systems
{\it Phys. Rev.} B  {\bf 32}  1778

\bibitem{LGK-91}
Lee J, Granato E and  Kosterlitz J M 1991
Nonuniversal critical behavior and first-order transitions
in a coupled XY-Ising model
{\it Phys. Rev.} B {\bf 44} 4819  

\bibitem{HPV-05let}
Hasenbusch M, Pelissetto A and Vicari E 2005
Transitions and crossover phenomena in fully frustrated XY systems
{\it Preprint} cond-mat/0506345

\bibitem{GKLN-91}
Granato E, Kosterlitz J M, Lee J and Nightingale M P 1991
Phase transitions in coupled XY-Ising systems
{\it Phys. Rev. Lett.} {\bf 66} 1090

\bibitem{HPV-05lungo}
Hasenbusch M, Pelissetto A and Vicari E 2005
The fully frustrated XY model and related systems,
in preparation

\bibitem{Kawamura-98}
Kawamura H 1998
Universality of phase transitions of frustrated antiferromagnets
\JPCM {\bf 10} 4707;
{\it Preprint} cond-mat/9805134

\bibitem{PV-review}
Pelissetto A and Vicari E 2002
Critical phenomena and renormalization-group theory
{\it Phys. Rept.} {\bf 368} 549;
{\it Preprint} cond-mat/0012164

\bibitem{francesi-review}
Delamotte B, Mouhanna D and Tissier M 2004
Nonperturbative renormalization-group approach to frustrated magnets
{\it Phys. Rev.} B {\bf 69} 134413

\bibitem{CPV-04}
Calabrese P, Parruccini P, Pelissetto A and Vicari E 2004
Critical behavior of O(2)$\otimes$O($N$)-symmetric models
{\it Phys. Rev.} B {\bf 70} 174439;
{\it Preprint} cond-mat/0405667

\bibitem{Olsson-95}
Olsson P 1995 
Two phase transitions in the fully frustrated XY model
{\it Phys. Rev. Lett.} {\bf 75} 2758

\bibitem{GPP-98}
Gro{\ss}e Pawig S and  Pinn K 1998
Monte Carlo algorithms for the fully frustrated XY model
{\it Int. J. Mod. Phys.} C {\bf 9} 727;
{\it Preprint} cond-mat/9807137

\bibitem{Amit-book}
Amit D J and Mart\'\i n-Mayor V 2005
{\it Field Theory, the Renormalization Group, and Critical Phenomena},
third ed. (Singapore: World Scientific)

\bibitem{LS-84} 
Lawrie D and Sarbach S 1984
Theory of tricritical points
{\em Phase Transitions and Critical Phenomena}
edited by C. Domb and J. L. Lebowitz
(London: Academic Press), Vol. 9

\bibitem{KNF-76}
Nelson D R, Kosterlitz J M  and Fisher M E 1974
Renormalization-group analysis of bicritical and tetracritical points
\PRL {\bf 33} 813

\bibitem{BDGL-86}
Berge B, Diep H T, Ghazali A and Lallemand P 1986
Phase transitions in two-dimensional uniformly frustrated XY spin systems
{\it Phys. Rev.} B {\bf 34} 3177

\bibitem{EHKT-89}
Eikmans H, van Himbergen J E, Knops H J F and Thijssen J M 1989
Critical behavior of an array of Josephson juctions
with variable couplings
{\it Phys. Rev.} B {\bf 39} 11759

\bibitem{GKS-98}
Granato E, Kosterlitz J M and  Simkin M V 1998
Edge effects in a frustrated Josephson-junction array
with modulated couplings
{\it Phys. Rev.} B {\bf 57} 3602;
{\it Preprint} cond-mat/9710242

\bibitem{CP-01}
Calabrese P and Parruccini P 2001
Critical behavior of two-dimensional frustrated spin models with 
noncollinear order
{\it Phys. Rev.} B {\bf 64} 184408;
{\it Preprint} cond-mat/0105551

\bibitem{PV-04}
Pelissetto A and Vicari E 2005
Interacting $N$-vector order parameters with $O(N)$ symmetry
{\it Cond. Matter Phys. (Ukraine)} {\bf 8} 87;
{\it Preprint} hep-th/0409214

\bibitem{Hasenbusch-99}
Hasenbusch M 1999 
A Monte Carlo study of leading order scaling corrections of 
$\phi^4$ theory on a three-dimensional lattice
{\it J. Phys.} A: {\it Math. Gen.} {\bf 32} 4851;
{\it Preprint} hep-lat/9902026

\bibitem{CHPRV-01}
Campostrini M, Hasenbusch M, Pelissetto A, Rossi P and Vicari E 2001
Critical behavior of the XY universality class
{\it Phys. Rev.} B {\bf 63} 214503;
{\it Preprint} cond-mat/0010360

\bibitem{SS-00}
Salas J and Sokal A D 2000 
Universal amplitude ratios in the critical
two-dimensional Ising model on a torus
{\it J. Stat. Phys.} {\bf 98} 551;
{\it Preprint} cond-mat/9904038

\bibitem{CHPV-02}
Caselle M, Hasenbusch M, Pelissetto A and Vicari E 2002
Irrelevant operators in the two-dimensional Ising model
\JPA {\bf 35} 4861; {\it Preprint} cond-mat/0106372

\bibitem{Hasenbusch-05}
Hasenbusch M 2005 
The two dimensional XY model at the transition temperature:
a high precision Monte Carlo study
{\it J. Phys.} A: {\it Math. Gen.} {\bf 38} 5869;
{\it Preprint} cond-mat/0502556

\end{thebibliography}
\end{document}